\documentclass[a4paper]{article}
\usepackage[margin=2.5cm]{geometry}
\usepackage{amsmath}
\usepackage{amssymb}
\usepackage{graphicx}
\usepackage{float}
\usepackage{hyperref}

\title{Marking Correlation}
\author{Paul Dubois, Romain Lhotte}

\newcommand{\Var}[1]{\text{Var}\left( {#1} \right)}

\begin{document}
	\maketitle
	\begin{abstract}
		Evaluating the performance of students in higher education is essential for gauging the effectiveness of teaching methods and achieving greater equality of opportunities for all.
		In this study, we investigate the correlation between two teachers' grading practices in a deep learning course at the master's level, offered at CentraleSupélec.
		The two teachers, who have distinct teaching styles, were responsible for marking the final project oral presentation.
		Our results indicate a significant positive correlation (0.76) between the two teachers' grading practices, suggesting that their assessments of students' performance are consistent.
		Although consistent with each other, grades do not seem to be fully reproducible from one examiner to the other suggesting serious drawbacks of only using one examiner for oral projects.
		Furthermore, we observed that the maximum difference between the grades assigned by the two examiners was 12.5\%, with a mean of 6.3\% (and median of 5.0\%), highlighting the potential impact of inter-examiner variability on students' final grades.
	\end{abstract}
	
	\section{Introduction}
	In recent years, there has been a surge of interest in the field of deep learning, with applications ranging from computer vision\cite{Voulodimos2018}\cite{CHAI2021} and natural language processing\cite{sun2021} to bio-informatics\cite{MIN2016}\cite{Zhang2020} and medical applications\cite{JAMIA2019}.
	As a result, there is an increasing demand for high-quality education and training programs in deep learning\cite{ResearchAndMarkets2023}.
	In response to this demand, many universities and engineering schools have started offering courses and programs in deep learning at various levels, including undergraduate and graduate levels.
	
	The evaluation of students' performance in the course is a critical aspect of teaching \cite{Karaman2011}\cite{Efu2019}.
	It can be used to measure the effectiveness of the teaching methods and the quality of the learning outcomes.
	In this study, we investigate the correlation between the grading practices of two instructors who taught the course, and the Kaggle assessments.
	The Kaggle assessments are regarded as impartial marking, and considered as a "ground truth".
	Our goal is to evaluate the reliability and consistency of grading practices in higher education and to provide insights that can inform efforts to improve the quality of teaching and learning outcomes.
	
	The findings of this study have implications for educators, administrators, and policymakers who are interested in improving the quality of education in deep learning and related fields.
	The results can also be used to inform the development of more effective evaluation frameworks and grading practices in higher education.
	Overall, this study contributes to the growing body of knowledge on the pedagogy of teaching scientific courses at advanced university level.
	
	\section{Methods}
	\subsection{Participants}
	The participants in this study were 28 students enrolled in a deep learning course at the master's level offered at CentraleSupélec during the academic year 2022/2023.
	The class was taught by two instructors, both of whom had distinct teaching styles.
	
	The course was designed to provide students with a comprehensive understanding of the fundamental concepts, theories, and applications of deep learning.
	The course was structured into 10 teaching sessions, each spanning three hours.
	Each session comprised one hour of theoretical instruction followed by two hours of practical exercises.
	In addition to the classroom instruction, five Kaggle challenges were assigned to the students, who were expected to work on them independently and outside of class time.
	The final component of the course consisted of a group project, which required all 28 students to form 10 groups of 2-4 individuals.
	The project served as a comprehensive assessment of the students' proficiency in deep learning and required the application of the concepts and techniques covered in the course.
	Students were expected to present their project findings orally and respond to questions from the instructors.

	\subsection{Evaluation Framework}
	The evaluation of students' performance in the course was based on six components, which accounted for the final grade.
	
	The first five components were Kaggle challenges, each accounting for 6\% of the total mark, adding up to 30\% of the final mark.
	For each challenge, a marking scheme was established, which was publicly shared with the students:
	\begin{itemize}
		\item [\textbf{\textit{0/6}}] \textbf{no attempt} to the challenge or score \textit{below} a \textit{basic} attempt of the challenge
		\item [\textbf{3/6}] achieved a better score than one typically obtained after a \textbf{basic} attempt of the challenge
		\item [\textbf{4/6}] achieved a better score than one typically obtained after a \textbf{fair} attempt of the challenge
		\item [\textbf{5/6}] achieved a better score than one typically obtained after an \textbf{advanced} attempt of the challenge
		\item [\textbf{6/6}] given to the \textbf{top 10} students of the class; to obtain 6/6, students also needed to meet requirement for 5/6 (the class is 28 students, so this score is reachable for a third of the class)

	\end{itemize}
	Details on the exact scores level for each Kaggle challenges are explained on the official website of the kaggles:
	\begin{itemize}
		\item [Kaggle 1] \url{https://www.kaggle.com/competitions/fitting-a-1d-1d-function-with-deep-learning}
		\item [Kaggle 2] \url{https://www.kaggle.com/competitions/fitting-a-5d-5d-function-with-deep-learning}
		\item [Kaggle 3] \url{https://www.kaggle.com/competitions/binary-classification-glasses-no-glasses}
		\item [Kaggle 4] \url{https://www.kaggle.com/competitions/decimal-classification-mics-mnist}
		\item [Kaggle 5] \url{https://www.kaggle.com/competitions/sword-video-classification}
	\end{itemize}
	
	The final component was a project, which accounted for the remaining 70\% of the final mark.
	Students completed the project in groups of 2-4, and the project was assessed based on an oral presentation of approximately 10-15 minutes, followed by approximately 10 minutes of questions.
	The final grade for the project was determined by averaging the independent scores provided by the two instructors, who evaluated the projects separately and without knowledge of each other's assigned scores.
	
	\subsection{Unbiased versus Biased Evaluation}
	The Kaggle challenges were employed as an impartial means of evaluating the students.
	The five Kaggle challenges were deemed to constitute a representative measure of the students' abilities.
	Teachers may harbor preferences for certain projects, be subject to halo effect or other types of biases\cite{Darby2007}\cite{Feeley2002}.
	As a result, the average score attained by the students in the Kaggle competitions was regarded as the "ground truth" for grading.
	It should be noted that certain students may have invested more time and effort in the project than in the Kaggle challenges, therefore making the Kaggle metric imperfect.
	Such factors, however, lie beyond the scope of this article.
	
	\subsection{Ethical Considerations}
	This study was conducted in accordance with the ethical guidelines of CentraleSupélec, and all participants were were anonymized for confidentiality purposes.
	
	\subsection{Normalizing}
	The Kaggles were originally graded out of 6, the project out of 20.
	We normalized all grades to be out of 100 (\%), to ba able to compare mean and standard deviation.
	
	\section{Results}
	We rounded all numerical values to $10^{-2}$ for readability purposes.
	\subsection{General results}
	We obtained the following results ($mean \pm standard\_deviation$):
	\begin{itemize}
		\item Kaggle 1: $88.89 \pm 9.25$
		\item Kaggle 2: $80.25 \pm 26.16$
		\item Kaggle 3: $89.51 \pm 8.20$
		\item Kaggle 4: $75.31 \pm 24.62$
		\item Kaggle 5: $70.06 \pm 31.37$
	\end{itemize}
	Kaggle Average: $80.80 \pm 13.22$
	\begin{itemize}
		\item Project-Teacher 1: $82.13 \pm 8.20$
		\item Project-Teacher 2: $80.83 \pm 11.22$
	\end{itemize}
	Project: $81.48 \pm 9.13$
	
	The Kaggle and project have similar average.
	The standard deviation is slightly lower on the project, but both are still comparable.
	
	\subsection{Dependence of assessments}
	Suppose that the score of students on each kaggle is independent.
	Then probability theory tells us that the standard deviation of the average kaggle score should be $9.85$ (using $\Var{\frac{\sum_{i=1}^{5} K_i}{5}} = \frac{\sum_{i=1}^{5} \Var{K_i}}{5^2}$).
	However, the variance found is $13.22$.
	This means that the score on kaggle challenges is not independent of others.
	It makes sense, since we expect good students to perform well in all assessments.
	
	Similarly, the variance of the project mark (average of the mark given by teacher 1 and 2) should be $6.95$ (using $\Var{\frac{P_1 + P_2}{2}} = \frac{\Var{P_1} + \Var{P_2}}{2^2}$) if the two teachers' grades could be treated as independent random variables.
	However, we observe a variance of $9.13$, leading to the conclusion that the two teachers marks cannot be treated as random independent variables.
	Again, this is expected, since teachers do not mark randomly.
	
	\subsection{Distribution of the marks}
	We can plot the Kernel Density Estimate (KDE) for the distribution of the marks, both for Kaggle challenges and for the project:
	\begin{figure}[H]
		\centering
		\includegraphics[width=0.49\textwidth]{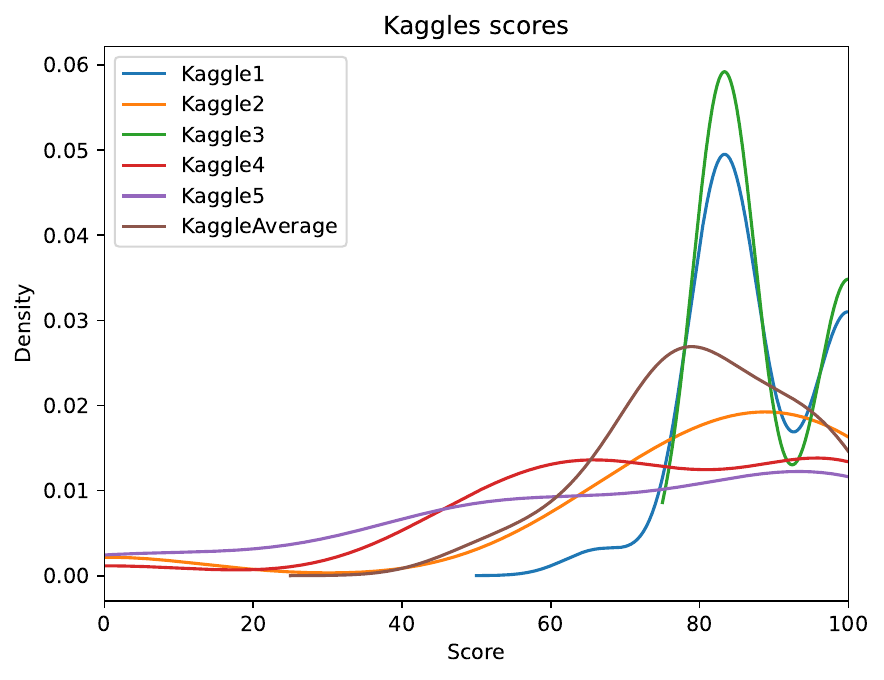}
		\includegraphics[width=0.49\textwidth]{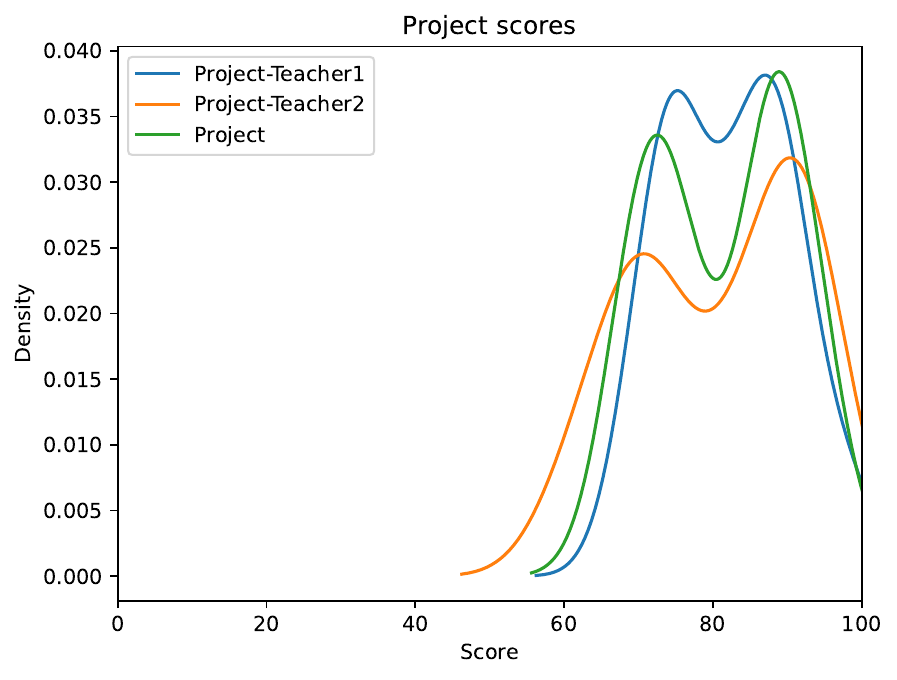}
		\caption{KDE}
		\label{fig:KDE}
	\end{figure}
	We observe that the class splits into two groups: one achieving nearly perfect score, and one getting about $75\%$.
	This trends can be observed in the project scores given by teachers 1 and 2, but also on some of the kaggle challenges: (kaggles 1 and 3 where it's obvious to observe, and kaggle 4 and 5 where it's less sharp).
	
	\subsection{Quantification of the inter-assessments correlation}
	You may find a complete table of correlation between all scores in the supplementary material.
	Here are the most interesting/interpretable ones:
	\begin{itemize}
		\item Kaggles-Project correlation: $0.64$; maximum absolute difference: $24.16\%$
		\item Inter-examiner correlation: $0.76$; maximum absolute difference: $12.50\%$
		\item Examiner-Kaggles correlation:\\
		Examiner 1: $0.49$\\
		Examiner 2: $0.69$
		\item Kaggle-Kaggle average correlation:\\
		Kaggle 1 - Kaggle Avg: $0.64$\\
		Kaggle 2 - Kaggle Avg: $0.53$\\
		Kaggle 3 - Kaggle Avg: $0.33$\\
		Kaggle 4 - Kaggle Avg: $0.71$\\
		Kaggle 5 - Kaggle Avg: $0.83$
	\end{itemize}

	\section{Discussion}
	The observed higher inter-examiner correlation as compared to the examiner-Kaggles correlation could potentially suggest that there are inherent differences in the oral presentation skills of the students being evaluated.
	While the examiners demonstrate a higher degree of agreement in their evaluations of the oral presentations, their scores do not align as closely with those assigned by the impartial non-oral assessment provided by the Kaggles.
	These findings highlight the potential limitations of using oral assessments for evaluating students.
	This suggests that there may be individual differences in the oral presentation proficiency of the students under evaluation.
	
	The observed low correlation between the Kaggles and their respective average scores suggests that students may have directed their attention towards particular Kaggles.
	This can be explained by two facts:
	First, students might have found some Kaggles more interesting than others, ans therefore performed better on those.
	Second, since there was an extra point for being in the top 10, students may have focused their attention on one or two Kaggle to get the extra point.
	This may have resulted in an uneven distribution of effort across the different Kaggles.
	The average of the Kaggle, however, is still a good unbiased metric.
	
	While the study provides valuable insights about how oral presentations may be appreciated differently by different examiners, it also has several limitations.
	
	Firstly, the study only included one class with 28 students, which means that the findings may not be generalizable to other settings or larger sample sizes.
	Secondly, the study only used one project assessment to evaluate student learning outcomes, which may not provide a comprehensive picture of student performance or achievement.
	Thirdly, the study only had two examiners, which may limit the reliability and validity of the assessment process.
	
	Additionally, it is possible that students may have paid more attention to one aspect of the course (either the Kaggle competition or the class project) than the other.
	This could create of decorrelation between the two marks that is not due to examiner's subjective appreciation, but to an objectively better/worst performance than in the Kaggle competitions.
	
	Further research with larger sample sizes and more diverse student populations is needed to confirm and extend these findings.

	\bibliographystyle{plain}
	\bibliography{references}

	\begin{center}
		\textbf{\LARGE Supplementary Materials}
	\end{center}

	\section*{Complete table of correlation between all scores}
	\begin{table}[!h]
		\begin{tabular}{|l|c|c|c|c|c|c|c|c|c|}
			\hline
			{} & Kaggle 1 & Kaggle 2 & Kaggle 3 & Kaggle 4 & Kaggle 5 & PT 1 & PT 2 & K Avg & PFM \\
			\hline
			Kaggle 1 & 1.00 & 0.56  & 0.23  & 0.30  & 0.30 & 0.09 & 0.36 & 0.64 & 0.26 \\
			Kaggle 2 & 0.56 & 1.00  & -0.01 & -0.01 & 0.12 & 0.07 & 0.36 & 0.53 & 0.25 \\
			Kaggle 3 & 0.23 & -0.01 & 1.00  & 0.10  & 0.29 & 0.18 & 0.29 & 0.33 & 0.26 \\
			Kaggle 4 & 0.30 & -0.01 & 0.10  & 1.00  & 0.61 & 0.56 & 0.47 & 0.71 & 0.54 \\
			Kaggle 5 & 0.30 & 0.12  & 0.29  & 0.61  & 1.00 & 0.46 & 0.59 & 0.83 & 0.57 \\
			PT1      & 0.09 & 0.07  & 0.18  & 0.56  & 0.46 & 1.00 & 0.76 & 0.49 & 0.92 \\
			PT2      & 0.36 & 0.36  & 0.29  & 0.47  & 0.59 & 0.76 & 1.00 & 0.69 & 0.96 \\
			K Avg    & 0.64 & 0.53  & 0.33  & 0.71  & 0.83 & 0.49 & 0.69 & 1.00 & 0.64 \\
			PFM      & 0.26 & 0.25  & 0.26  & 0.54  & 0.57 & 0.92 & 0.96 & 0.64 & 1.00 \\
			\hline
		\end{tabular}
		\caption{Correlation between scores obtained by students.}
		\label{table:all_correlations}
	\end{table}
	\vspace{0.1cm}
	\textit{(All values have been rounded to $10^{-2}$.)}

	\vspace{0.5cm}
	Abbreviation :
	\begin{itemize}
		\item PT 1: Project Teacher 1's mark
		\item PT 2: Project Teacher 2's mark
		\item K Avg: Kaggle Average
		\item PFM: Project Final Mark
	\end{itemize}
	
\end{document}